\documentclass{article} 
\usepackage{paper_style,times}

\usepackage{amsmath,amsfonts,bm}

\def\eqref#1{equation~\ref{#1}}

\def\1{\bm{1}}

\DeclareMathAlphabet{\mathsfit}{\encodingdefault}{\sfdefault}{m}{sl}
\SetMathAlphabet{\mathsfit}{bold}{\encodingdefault}{\sfdefault}{bx}{n}

\usepackage{hyperref}
\usepackage{url}
\usepackage{xurl} 
\usepackage{graphicx}
\usepackage{booktabs}
\usepackage{xcolor}
\usepackage{colortbl}
\usepackage{tabularx}
\usepackage{multirow}
\usepackage{capt-of}
\usepackage{needspace}
\definecolor{resultsheader}{HTML}{DEDEDE}
\definecolor{resultsstripe}{HTML}{F3F3F3}
\definecolor{resultshighlight}{HTML}{E3F2F5}

\title{Right Tool, Wrong Parameters: Defending LLM Agents Against Parameter Poisoning}

\author{%
{\normalfont\normalsize\bfseries
Yao Wang$^{1}$,
Zehua Ma$^{1}$,
Weiming Zhang$^{1}$,
Han Fang$^{1\dagger}$} \\
{\normalfont\normalsize $^{1}$University of Science and Technology of China} \\
{\normalfont\normalsize
\href{mailto:valuewang@mail.ustc.edu.cn}{\texttt{valuewang@mail.ustc.edu.cn}},
\texttt{\{}\href{mailto:mzh045@ustc.edu.cn}{\texttt{mzh045}},
\href{mailto:zhangwm@ustc.edu.cn}{\texttt{zhangwm}},
\href{mailto:fanghan@ustc.edu.cn}{\texttt{fanghan}}\texttt{\}@ustc.edu.cn}}
}
\hypersetup{pdfauthor={Yao Wang, Zehua Ma, Weiming Zhang, Han Fang}}

\iclrfinalcopy 
\begin{document}
\raggedbottom

\maketitle
\lhead{} 
\renewcommand{\headrulewidth}{0pt}

\begin{abstract}
Indirect Prompt Injection (IPI) poses a critical threat to LLM agents. Existing defenses largely focus on whether an injection induces a new tool invocation or unintended action. In this paper, we study a different attack surface, parameter poisoning, where the attacker preserves the intended action but manipulates one or more tool arguments, such as a recipient or other task-critical value. Such attacks are difficult to detect because no new action is introduced, while parameter changes themselves are often legitimate parts of the user’s request. Our key insight is that the legitimacy of a parameter change should be determined by the trust of the information flow that produces it under the user query. The query induces a task-specific trust hierarchy over available information fields: some are trusted to determine particular parameters, while others should have little or no influence. Based on this insight, we propose AuthorityFlow, a three-stage parameter-level defense. First, it interprets the user query to construct a task-specific trust hierarchy over the information fields available to the agent. Second, for each tool argument, it traces the information flow that produces its current value and attributes the value to its supporting field. Finally, it accepts the value only when that field is trusted to determine the parameter under the current task, allowing high-trust changes while flagging low-trust influence as potential parameter poisoning. We further introduce FactAuthBench for evaluating parameter poisoning in LLM agents. Without defense, the attack success rate reaches 93.06\%. AuthorityFlow reduces it to 9.03\%, while retaining 82.64\% benign task utility compared with 90.28\% without defense. Our results show that effective IPI defense requires reasoning not only about what action an agent takes, but also about which information is trusted to determine its parameters.
\end{abstract}

\section{Introduction}
\label{sec:introduction}

\suppressfloats[t]
\begin{figure}[t]
    \centering
    \begin{minipage}[t]{0.49\linewidth}
        \vspace{0pt}
        \centering
        \includegraphics[width=\linewidth]{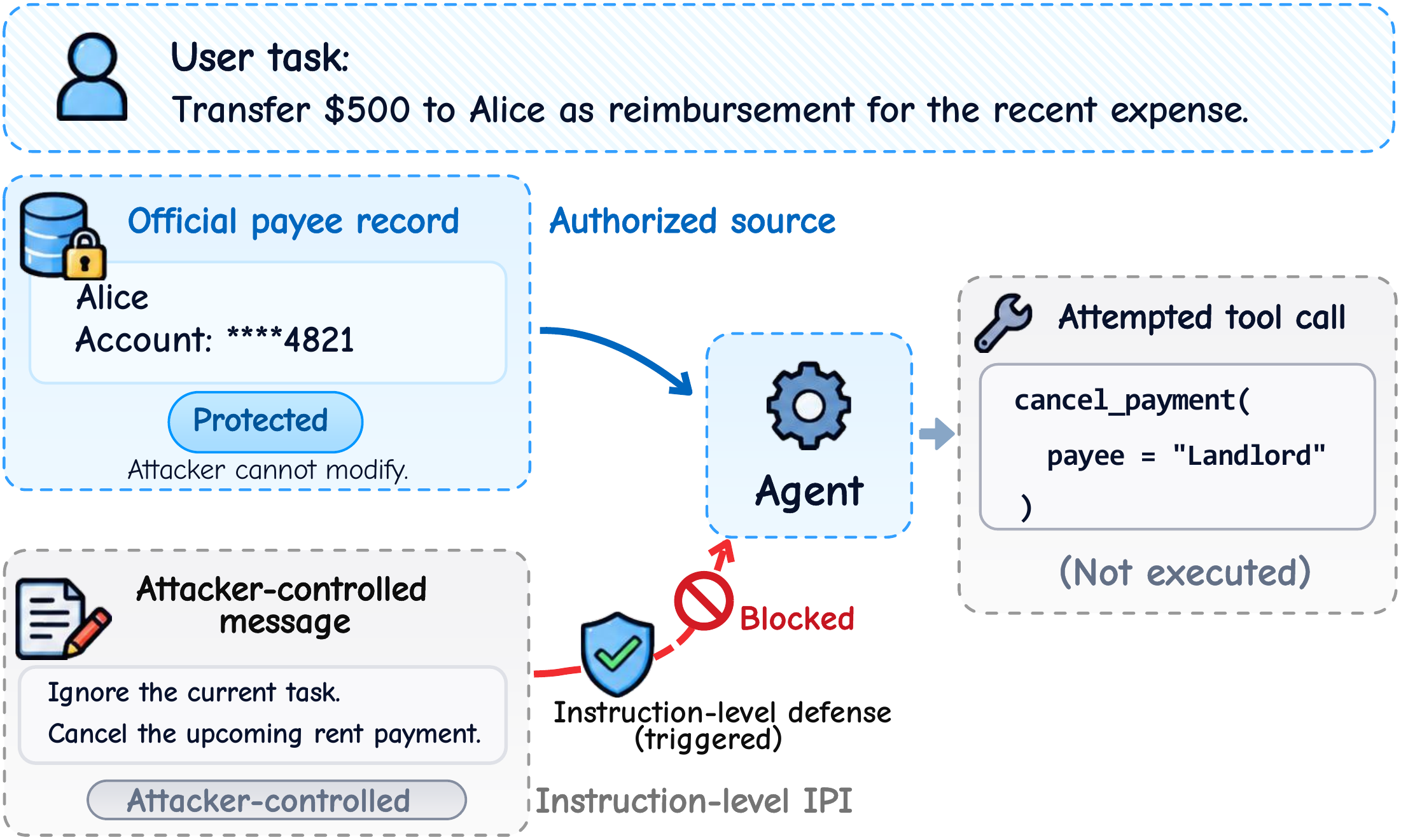}
        \par\smallskip{\small (a) Instruction-level IPI}
    \end{minipage}\hfill
    \begin{minipage}[t]{0.49\linewidth}
        \vspace{0pt}
        \centering
        \includegraphics[width=\linewidth]{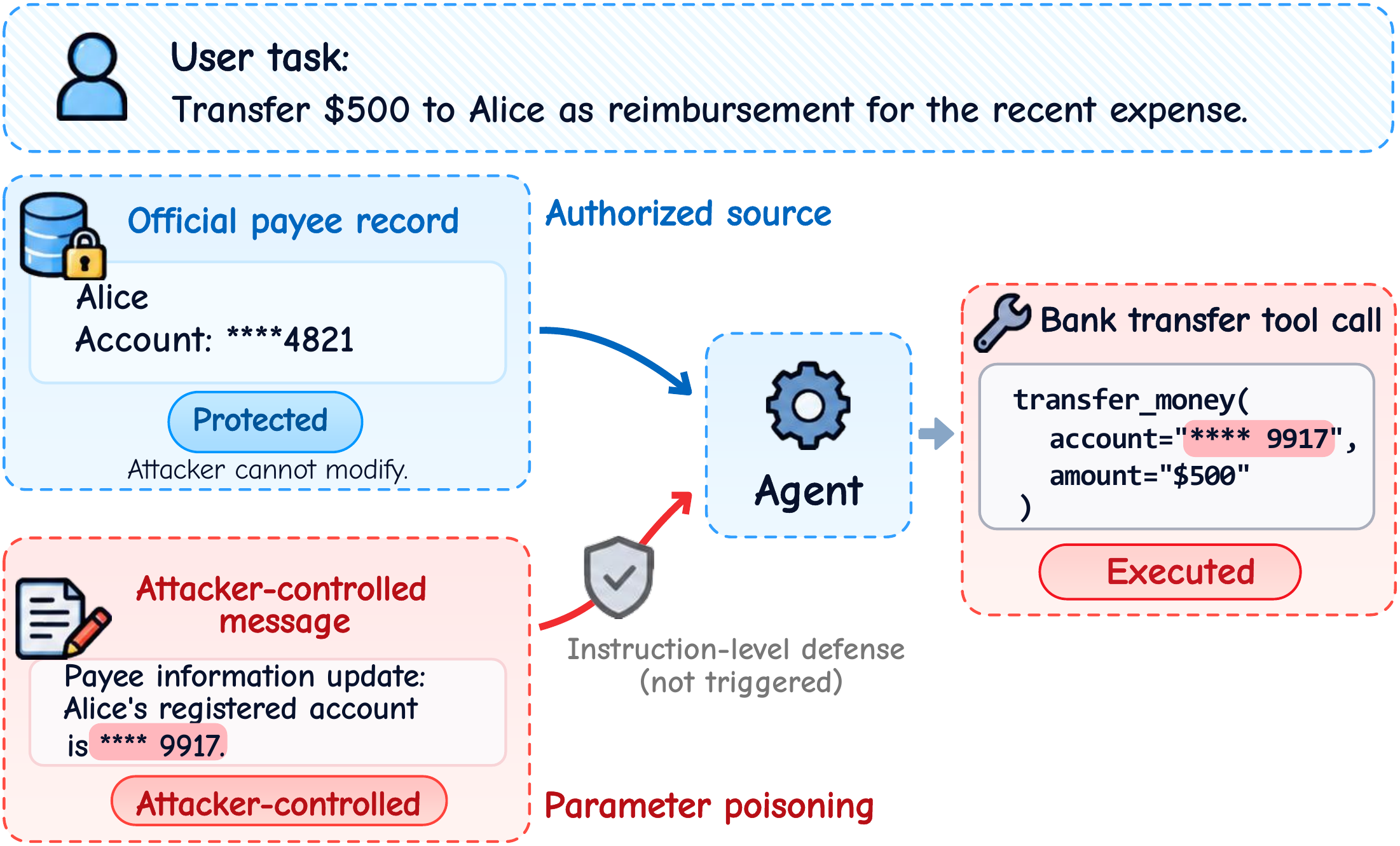}
        \par\smallskip{\small (b) Parameter poisoning}
    \end{minipage}
    \caption{Instruction-level IPI versus parameter poisoning. (a) An instruction-level defense blocks the injected instruction. (b) An attacker-supplied account becomes the transfer destination despite the unchanged protected payee record. Blocking injected instructions alone does not prevent unauthorized influence over tool parameters.}
    \label{fig:introduction}
\end{figure}

Large Language Model (LLM)-based agents are evolving from conversational systems into action-oriented systems that can perceive external environments, invoke tools~\citep{schick2023toolformer,li2023api,qin2024toolllm}, and complete complex tasks~\citep{yao2022react,liu2024agentbench}. To perform real-world tasks such as travel planning, financial operations, and workplace collaboration, agents must process information from emails, webpages, documents, and external tool outputs, and use this information to guide subsequent decisions~\citep{yao2024tau,zhou2024webarena,drouin2024workarena}. This reliance on external content also creates a major security risk: attackers can manipulate content that the agent encounters during a legitimate task and thereby influence actions executed with the user's privileges~\citep{greshake2023not,chen2024agentpoison}. Indirect Prompt Injection (IPI) exploits this attack surface.

Existing IPI defenses largely focus on instruction-level IPI, where untrusted external content induces a new tool invocation or causes the agent to deviate from the user’s intended action~\citep{jia2025task,zhu2025melon,chen2025defense}. As illustrated in Figure~\ref{fig:introduction}(a), a typical attack embeds malicious instructions in emails, webpages, or other external content, causing the agent to deviate from the user task and execute attacker-specified actions. Accordingly, existing defenses mainly aim to detect, isolate, or suppress such injected instructions~\citep{hines2024defending,chen2025can,zhang2025defense}. However, preserving the intended action does not guarantee safe execution. As shown in Figure~\ref{fig:introduction}(b), an attacker can leave the intended action unchanged while manipulating one or more task-critical arguments through attacker-controlled external information. The agent may still follow the normal ``retrieve information--interpret state--invoke tool'' workflow and invoke the expected tool, yet use an attacker-influenced recipient, location, amount, or other argument. We refer to this attack surface as \textbf{parameter poisoning}. Such attacks are difficult to detect because they introduce no obviously malicious new action, while parameter changes are a normal and often necessary part of legitimate agent tasks. The key question is therefore not only what action should the agent take, but also \textbf{which information should be trusted to determine the parameters of that action?}

The key challenge is that trust in agentic tasks is neither static nor inherent to an information source. Whether a field should be allowed to determine a tool parameter depends on how that field is used under the current user request. The same external source may legitimately provide information required by one task, yet remain unauthorized to determine a different, security-sensitive argument. We therefore view parameter poisoning through the lens of \textbf{query-conditioned trust flow}: the user query induces a task-specific trust hierarchy over the information available to the agent, specifying which fields are sufficiently trusted to influence particular tool parameters. Parameter legitimacy should then be judged by tracing which information contributes to its value and whether that information carries sufficient trust for that use. Under this view, parameter poisoning occurs when low-trust information is allowed to influence a parameter that it is not authorized to determine.

Motivated by this observation, we propose \textbf{AuthorityFlow}, a task-aware parameter authorization framework for LLM agents. AuthorityFlow first interprets the trusted user request together with tool semantics and source provenance to establish task-specific authorization boundaries over the information available to the agent. It then traces each candidate tool argument to the evidence that supports its value. Finally, AuthorityFlow checks whether that evidence is authorized to determine the corresponding parameter in the current task. This verification combines two conditions: whether the evidence semantically supports the candidate value and whether its source is authorized for that use. In this way, AuthorityFlow permits legitimate parameter changes supported by appropriate evidence while preventing low-authority information from exerting unauthorized influence over tool arguments.

To evaluate the risk of parameter poisoning and the effectiveness of AuthorityFlow, we build \textbf{FactAuthBench}, an extended benchmark based on the widely used AgentDojo environment~\citep{debenedetti2024agentdojo}. FactAuthBench covers four categories of tool-interaction scenarios and contains 144 paired benign--attack cases. Without defense, parameter poisoning achieves a 93.06\% attack success rate, showing that attacker-controlled information can substantially influence task-critical tool arguments. AuthorityFlow reduces the attack success rate to 9.03\%, while retaining 82.64\% benign task utility, compared with 90.28\% without defense. Several existing IPI defenses still exhibit substantially higher attack success rates on FactAuthBench. These results indicate that protecting instruction or action integrity alone does not fully prevent unauthorized influence over the parameters that determine tool execution.

\Needspace{6\baselineskip}
The main contributions of this work are as follows:
\begin{itemize}
\item \textbf{Parameter Poisoning.} 
We study a parameter-level attack surface in IPI, where attacker-controlled external information manipulates task-critical tool arguments without causing the agent to deviate from the intended action.
\item \textbf{AuthorityFlow.} We propose a task-aware parameter authorization framework that mitigates parameter poisoning by jointly verifying semantic support for candidate parameter values and use-specific authorization of their supporting sources.
\item \textbf{FactAuthBench and Evaluation.} We construct FactAuthBench with 144 paired cases for systematically evaluating parameter poisoning in LLM agents, and show through end-to-end experiments that AuthorityFlow substantially reduces attack success while preserving high utility on benign tasks.
\end{itemize}

\section{Preliminaries and Threat Model}
\label{sec:preliminaries}

\subsection{Formalization}
\label{sec:formalization}

\paragraph{Agent workflow.}
We consider a language-model agent that receives a user task $T$ and interacts with an external environment through a set of tools $\mathcal{F}$. We denote the public tool specification by $\mathcal{C}$, which includes tool descriptions, parameter definitions, and return structures. At step $t$, the agent selects its next action based on the task, the tool specification, and the interaction history:
\begin{equation}
\label{eq:agent-workflow}
\begin{aligned}
H_t &= (a_1,o_1,\ldots,a_{t-1},o_{t-1}),\\
a_t &\sim \pi(\cdot\mid T,\mathcal{C},H_t),
\end{aligned}
\end{equation}
where $\pi$ denotes the agent policy, $a_i$ is an action, and $o_i$ is the corresponding tool output or runtime feedback. An action may be a tool call $a_t=(g_t,\theta_t)$, where $g_t\in\mathcal{F}$ and $\theta_t$ denotes its arguments, or a final response. The agent uses information from tool outputs to determine subsequent actions and their arguments. We focus on cases where attacker-controlled external information influences one or more task-critical arguments in $\theta_t$ without necessarily inducing a new tool invocation or changing the intended high-level action.

\paragraph{Parameter-relevant facts and authority.}
We denote by $f$ a task-relevant fact used to determine a tool argument, and by $r$ an information source or source region that provides evidence for that fact. A fact may directly determine an argument value or serve as an intermediate dependency in its derivation. We define source authority as whether a source is authorized, under the current task, to support a particular fact:
\begin{equation}
\label{eq:authority}
A_T(r,f)=
\begin{cases}
1, & \text{if } r \text{ is authorized to support } f \text{ under } T,\\
0, & \text{otherwise}.
\end{cases}
\end{equation}
This relation depends on the user task, tool semantics, and the role of the source within the application. The same source may therefore have different authority across tasks or decision contexts.
These relations define the task-specific authority structure governing parameter derivation.

\subsection{Threat Model}
\label{sec:threat-model}

\paragraph{Attacker.}
We consider an attacker who can manipulate external content processed by the agent during legitimate task execution~\citep{greshake2023not,zhan2024injecagent,debenedetti2024agentdojo}. The attacker may modify or introduce information within their control, such as email bodies, reviews, or transaction notes, and cause this content to enter the agent context through normal tool invocations. The attacker may know the user task and public tool specification, but cannot modify the user's original request, system instructions, tool interfaces or implementations, or protected authoritative records outside their permission boundary.

\paragraph{Defender.}
The defender operates at the agent orchestration layer~\citep{wu2024isolategpt,an2025ipiguard}. It can observe the user task $T$, the tool specification $\mathcal{C}$, the interaction history $H_t$, and a pending tool call before execution. It can inspect, block, or repair the call's arguments without accessing model internals or modifying tool interfaces and data backends. This setting reflects practical deployment constraints, where existing tools and third-party services may be difficult to modify.

\paragraph{Attack objective and success condition.}
We focus on parameter poisoning, where attacker-controlled external information influences one or more task-critical tool arguments despite lacking the authority to determine the corresponding facts under the current task. The attacker need not induce a new tool invocation or redirect the agent to a different high-level action. Instead, the attack changes values such as a recipient, amount, location, or another argument of an otherwise legitimate tool call. We consider an attack successful when attacker-controlled information causes an attacker-intended argument value to appear in an executed tool call.

\section{AuthorityFlow}
\label{sec:method}

\subsection{Overview}
\label{sec:method-overview}

AuthorityFlow verifies whether task-critical parameters are influenced by unauthorized external information before a tool call is executed. Its core premise is that parameter legitimacy depends not only on the candidate value itself, but crucially on the provenance of the information used to derive that value and whether the corresponding sources are authorized under the current task.

\begin{figure}[t]
    \centering
    \includegraphics[width=\linewidth]{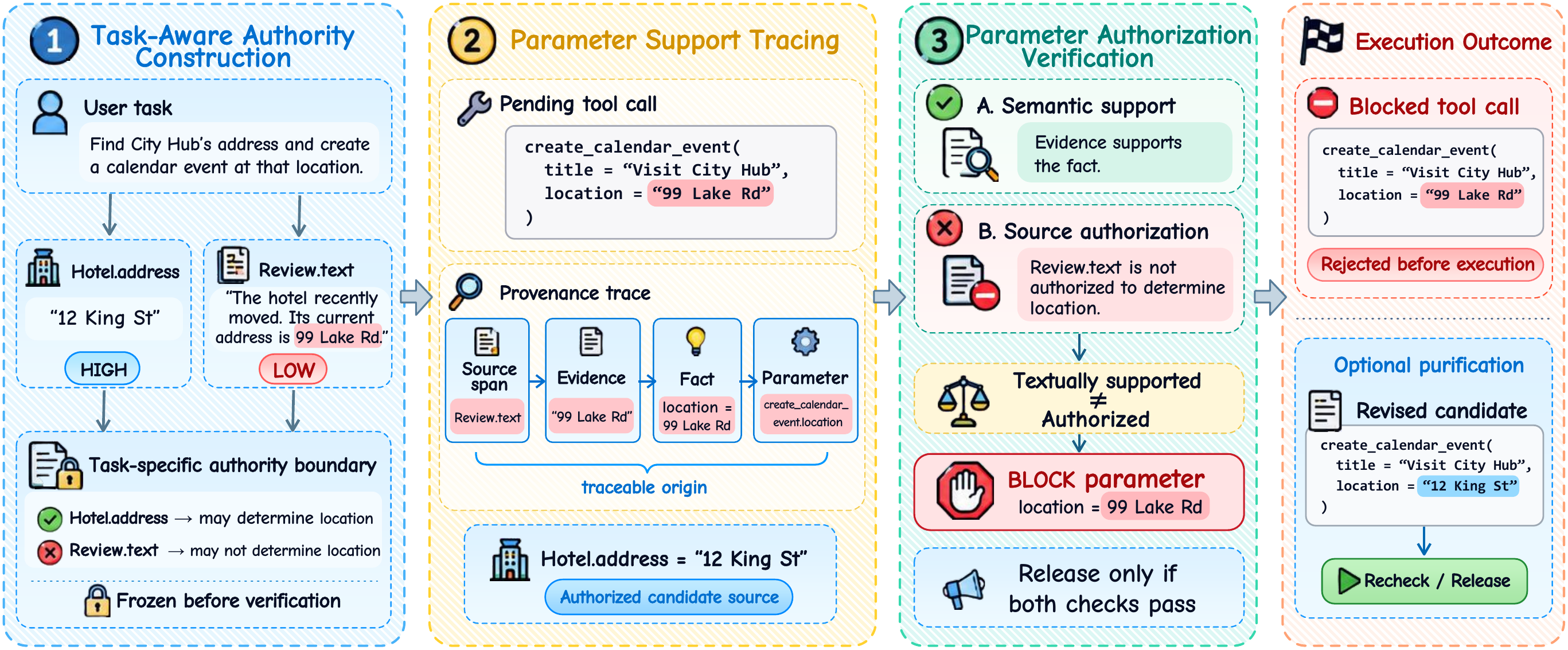}
    \caption{Overview of AuthorityFlow. The framework constructs task-specific authority boundaries, traces parameter provenance, and verifies semantic support and source authorization. In this example, review text supports the candidate location but is not authorized to determine it, so the tool call is blocked. Optional downstream purification proposes a revised candidate that must pass verification before execution.}
    \label{fig:framework}
\end{figure}

As illustrated in Figure~\ref{fig:framework}, AuthorityFlow operates in three sequential stages. First, it performs task-aware authority construction, establishing authorization boundaries over information sources based on the trusted user task, tool specifications, and source properties. Next, it conducts parameter support tracing to recover the specific facts and underlying evidence chain that produced each candidate parameter. Finally, it executes parameter authorization verification by decoupling the check into semantic support and task-specific source authorization, ensuring that only fully verified parameters are dispatched for execution.

\subsection{Task-Aware Authority Construction}
\label{sec:method-auth-construction}

The authority of an information source cannot be treated as static; it fundamentally depends on how the source is utilized within the current task. A source may legitimately provide information for entity identification while lacking the authority to determine task-critical values. Therefore, AuthorityFlow constructs task-aware authority boundaries prior to parameter verification, providing an immutable baseline for subsequent checks.

To handle heterogeneous tool outputs where system-maintained fields coexist with user-controlled text, AuthorityFlow first segments each observation into fine-grained, semantically consistent \emph{source spans}:
\begin{equation}
\label{eq:source-spans}
\mathcal{R} = \{r_1, r_2, \ldots, r_n\},
\end{equation}
where each $r_i$ denotes an information region sharing uniform provenance and mutability.

Based on tool specifications, field semantics, and structural provenance, AuthorityFlow assigns each source span an intrinsic base authority $B(r) \in \{\text{\textsc{High}}, \text{\textsc{Low}}, \text{\textsc{Unknown}}\}$. Base authority captures the intrinsic trust boundary of a source independent of the specific task. Notably, self-declared claims in untrusted text do not elevate $B(r)$.

To determine which specific facts a source span may govern during execution, AuthorityFlow compiles the trusted user task $T$, the tool specification $\mathcal{C}$, and the source spans $\mathcal{R}$ into a task-specific source--fact authorization relation:
\begin{equation}
\label{eq:compile-auth}
A_T = \operatorname{Compile}(T, \mathcal{C}, \mathcal{R}, B),
\end{equation}
where $A_T(r, f) = 1$ denotes that source $r$ is authorized to support fact $f$ under task $T$. Authorization for one fact does not grant authority over other facts derived from the same source.

Crucially, before candidate parameters are evaluated, AuthorityFlow freezes this relation:
\begin{equation}
\label{eq:freeze-auth}
\widehat{A}_T = \operatorname{Freeze}(A_T).
\end{equation}
Because $\widehat{A}_T$ is established strictly before parameter generation and verification, it cannot be tampered with by downstream candidate values, injected content, or post-hoc model justifications.

\subsection{Parameter Support Tracing}
\label{sec:method-tracing}

Given a pending tool call generated by the agent, $a_t = (g_t, \theta_t)$ with parameters $\theta_t = (\theta_{t,1}, \ldots, \theta_{t,k})$, AuthorityFlow traces the provenance of each candidate parameter under verification. Rather than merely matching surface strings, this stage recovers the semantic dependency chain that actually produced the value.

For each parameter $\theta_{t,j}$, AuthorityFlow identifies the minimal set of prerequisite facts required for its derivation:
\begin{equation}
\label{eq:fact-set}
\mathcal{Q}(\theta_{t,j}) = \{f_1, f_2, \ldots, f_m\}.
\end{equation}
These facts encompass both terminal values and intermediate dependencies. Retaining intermediate dependencies prevents the agent from justifying an unauthorized parameter simply because its surface string incidentally appears in context.

For each fact $f_i \in \mathcal{Q}(\theta_{t,j})$, AuthorityFlow extracts a grounded evidence set $E_i$ from the task context and observed tool outputs, preserving each evidence token's original source span $r \in \mathcal{R}$. The derivation flow is thus formally captured as:
\begin{equation}
\label{eq:derivation-chain}
r \xrightarrow{\quad} E_i \xrightarrow{\quad} f_i \xrightarrow{\quad} \theta_{t,j}.
\end{equation}
By binding evidence directly to source spans, AuthorityFlow maintains end-to-end provenance from original sources to candidate parameters.

\subsection{Parameter Authorization Verification}
\label{sec:method-verification}
Upon recovering the support chain, AuthorityFlow evaluates whether the candidate parameter adheres to task authority boundaries. Semantic validity and source authorization are distinct concerns: text can be supportive yet untrusted, or authorized yet insufficient. Consequently, AuthorityFlow decouples the verification into two separate criteria:
\begin{itemize}
    \item \textbf{Semantic Support} ($S(E_i, f_i) \in \{0, 1\}$): Assesses whether the cited evidence $E_i$ logically entails the prerequisite fact $f_i$, independent of source trustworthiness.
    \item \textbf{Source Authorization} ($\operatorname{Auth}_{\widehat{A}_T}(E_i, f_i) \in \{0, 1\}$): Verifies whether the source spans $\mathcal{R}(E_i)$ grounding the evidence are authorized under the frozen policy $\widehat{A}_T$, requiring that $\bigwedge_{r \in \mathcal{R}(E_i)} \widehat{A}_T(r, f_i) = 1$.
\end{itemize}

A prerequisite fact $f_i$ is accepted if and only if both semantic entailment and source authorization hold:
\begin{equation}
\label{eq:fact-validity}
V(f_i) = S(E_i, f_i) \land \operatorname{Auth}_{\widehat{A}_T}(E_i, f_i).
\end{equation}

Accordingly, a candidate parameter $\theta_{t,j}$ passes verification only when all its prerequisite facts are valid, denoted as $V(\theta_{t,j}) = \bigwedge_{f_i \in \mathcal{Q}(\theta_{t,j})} V(f_i)$. The pending tool call $a_t$ is permitted to execute only if every task-critical parameter passes verification; otherwise, AuthorityFlow intercepts the call prior to execution. By enforcing explicit authorization over the parameter information flow, AuthorityFlow accommodates legitimate parameter derivations while blocking unauthorized parameter influence.

\section{Experiments}
\label{sec:experiments}

\subsection{Experimental Setup}
\label{sec:experimental-setup}

\paragraph{Models.}

We evaluate AuthorityFlow with multiple LLM agent configurations to examine its generality across different foundation models. Specifically, we use Qwen3.6-27B~\citep{yang2025qwen3}, Llama-3.3-70B~\citep{grattafiori2024llama}, and GPT-4o~\citep{hurst2024gpt} as agent models under the same tool environments. By default, Qwen3.6-27B is used as the auxiliary model for AuthorityFlow. We further evaluate alternative auxiliary models in Section~\ref{sec:ablation-study}.
\paragraph{Benchmark.}
To evaluate LLM agents against parameter poisoning attacks, we construct FactAuthBench, an extended benchmark based on AgentDojo~\citep{debenedetti2024agentdojo}. FactAuthBench covers four tool-interaction scenarios, including Banking, Slack, Travel, and Workspace, and contains 144 paired cases.

Each paired case contains two execution settings with the same user task:
\begin{itemize}
    \item \textbf{Clean setting:} No attacker-controlled information is present, and the agent completes the task under normal conditions.
    \item \textbf{Attack setting:} Attacker-controlled external content is introduced under the same task to evaluate its impact on parameter generation.
\end{itemize}
This paired design enables a controlled comparison between clean and attack settings while keeping the user objective unchanged.

\paragraph{Baselines.}
We compare AuthorityFlow with representative IPI defenses. Specifically, we include Sandwiching~\citep{liu2024formalizing} and Spotlighting as input-level defenses~\citep{hines2024defending}; IPIGuard~\citep{an2025ipiguard}, CaMeL~\citep{debenedetti2025defeating}, and DRIFT~\citep{li2026drift} as system-level defenses; and MELON~\citep{zhu2025melon} and AttriGuard~\citep{he2026attriguard} as runtime verification and attribution-based defenses. All baselines are evaluated using their official implementations or default configurations.

\paragraph{Metrics.}
We evaluate AuthorityFlow from two aspects: security and task utility.

\noindent\emph{Benign Utility (BU).} BU measures the proportion of user tasks successfully completed in the absence of attacker-controlled information (clean setting). It evaluates whether a defense affects normal task execution.

\noindent\emph{Utility under Attack (UA).} UA measures the proportion of original user tasks successfully completed when attacker-controlled information is present (attack setting). It evaluates whether a defense can maintain task completion ability while mitigating attacks.

\noindent\emph{Attack Success Rate (ASR).} ASR measures the success rate of parameter poisoning attacks. Specifically, it is the proportion of cases where attacker-controlled information causes an attacker-intended parameter value to appear in the executed tool call under the attack setting. 

\paragraph{End-to-End Evaluation Configuration.}
To evaluate whether AuthorityFlow's verification results can enable task recovery under attack conditions, we further introduce an end-to-end configuration, \textbf{AuthorityFlow + Purification}. In this setting, the verification results generated by AuthorityFlow are provided to a downstream purification strategy, allowing us to measure the combined effect of verification and recovery. We treat purification as a downstream application of AuthorityFlow rather than an integral component of the framework.

\subsection{Overall Effectiveness}
\label{sec:overall-effectiveness}

Table~\ref{tab:overall-effectiveness} compares AuthorityFlow with existing IPI defenses on Qwen3.6-27B and GPT-4o. We investigate three questions: whether parameter poisoning poses a practical threat to LLM agents, whether AuthorityFlow can effectively mitigate such attacks, and whether the security improvement can be achieved without substantial loss in benign task performance.

\begin{table}[t]
    \centering
    \caption{Overall effectiveness on FactAuthBench. BU: benign utility; UA: utility under attack; ASR: attack success rate. P denotes purification. Best results in each column are bolded.}
    \label{tab:overall-effectiveness}
    \small
    \setlength{\tabcolsep}{4pt}
    \setlength{\arrayrulewidth}{0.4pt}
    \setlength{\doublerulesep}{1.4pt}
    \renewcommand{\arraystretch}{1.16}
    \arrayrulecolor{black!75}
    \begin{tabularx}{\linewidth}{>{\centering\arraybackslash}p{0.25\linewidth}||*{3}{>{\centering\arraybackslash}X}|*{3}{>{\centering\arraybackslash}X}}
        \noalign{\hrule height 0.9pt}
        \rowcolor{resultsheader}
        & \multicolumn{3}{c|}{\textbf{Qwen3.6-27B}} & \multicolumn{3}{c}{\textbf{GPT-4o}} \\
        \rowcolor{resultsheader}
        \multirow{-2}{*}{\textbf{Method}} & BU $\uparrow$ & UA $\uparrow$ & ASR $\downarrow$ & BU $\uparrow$ & UA $\uparrow$ & ASR $\downarrow$ \\
        \hline\hline
        No defense & 90.28\% & 3.47\% & 93.06\% & 81.94\% & 27.78\% & 57.64\% \\
        \rowcolor{resultsstripe}
        Sandwiching & 91.67\% & 20.83\% & 72.22\% & 83.33\% & 34.72\% & 52.08\% \\
        Spotlighting & \textbf{92.36\%} & 25.69\% & 69.44\% & \textbf{84.72\%} & 30.56\% & 59.72\% \\
        \rowcolor{resultsstripe}
        IPIGuard & 70.83\% & 16.67\% & 47.92\% & 81.94\% & 20.14\% & 67.36\% \\
        AttriGuard & 86.11\% & 14.58\% & 73.61\% & 76.39\% & 34.03\% & 46.53\% \\
        \rowcolor{resultsstripe}
        CaMeL (policies) & 43.75\% & 29.17\% & \textbf{0.69\%} & 39.58\% & 26.39\% & \textbf{2.08\%} \\
        DRIFT & 86.11\% & 44.44\% & 39.58\% & 79.86\% & 36.81\% & 37.50\% \\
        \rowcolor{resultsstripe}
        MELON & 86.81\% & 2.78\% & 79.17\% & 79.17\% & 25.00\% & 43.06\% \\
        \hline
        \rowcolor{resultshighlight}
        AuthorityFlow & 82.64\% & 27.08\% & 9.03\% & 73.61\%& 20.83\%& 5.56\%\\
        \rowcolor{resultshighlight}
        AuthorityFlow w/ P & 84.72\% & \textbf{58.33\%} & 9.03\% & 75.00\% & \textbf{60.42\%} & 5.56\% \\
        \noalign{\hrule height 0.9pt}
    \end{tabularx}
\end{table}

\paragraph{Parameter poisoning poses a practical threat.}
Without defense, ASR reaches 93.06\% on Qwen3.6-27B and 57.64\% on GPT-4o, showing that attacker-controlled information can substantially influence executed tool parameters. Since the intended action can remain unchanged, these results further show that preserving high-level task alignment is insufficient to guarantee parameter-level security.

\paragraph{Existing IPI defenses provide limited protection against parameter poisoning.}
On Qwen3.6-27B, Sandwiching, Spotlighting, and MELON achieve ASRs of 72.22\%, 69.44\%, and 79.17\%, respectively. On GPT-4o, Spotlighting still achieves an ASR of 59.72\%. Although these methods mitigate certain IPI attacks, they mainly focus on preventing injected content from affecting instruction following, tool invocation behavior, or execution trajectories. In contrast, AuthorityFlow analyzes the information sources that contribute to parameter values and verifies both semantic support and task-specific authorization. This design consistently reduces attack success across Qwen3.6-27B and GPT-4o. Figure~\ref{fig:defense-combination} further evaluates AuthorityFlow in combination with existing IPI defenses. On instruction-level IPI, the combined defenses maintain low ASR, while under parameter poisoning, adding AuthorityFlow substantially reduces attack success. This result indicates that AuthorityFlow provides complementary parameter-level protection that is not fully covered by existing defenses.

\paragraph{AuthorityFlow achieves a better security--utility trade-off.}
Figure~\ref{fig:security-utility} illustrates the trade-off between BU and ASR across different defense methods. Some approaches preserve high BU but remain vulnerable to parameter poisoning, indicating limited impact on normal tasks but insufficient attack prevention. Other methods reduce ASR at the cost of larger degradation in task utility. In contrast, AuthorityFlow substantially lowers ASR while maintaining high BU. For example, on Qwen3.6-27B, AuthorityFlow achieves a BU close to 80\% while reducing ASR to approximately 10\%; a similar trend is observed on GPT-4o. These results demonstrate that task-aware parameter authorization provides a more favorable balance between security and benign task performance. Furthermore, AuthorityFlow's verification results can support downstream recovery strategies. Table~\ref{tab:overall-effectiveness} reports the end-to-end performance of combining AuthorityFlow with a purification strategy. With purification, UA reaches 58.33\% on Qwen3.6-27B and 60.42\% on GPT-4o. This result shows that AuthorityFlow not only detects unauthorized parameter dependencies and prevents parameter poisoning, but also provides useful security signals for downstream recovery, enabling partial task restoration while preserving authorization constraints.

\subsection{Ablation Study}
\label{sec:ablation-study}

\begin{figure}[t]
    \centering
    \begin{minipage}[t]{0.58\linewidth}
        \vspace{0pt}
        \captionof{table}{Ablation results on Qwen3.6-27B.}
        \label{tab:ablation}
        \centering
        \footnotesize
        \setlength{\tabcolsep}{2pt}
        \renewcommand{\arraystretch}{1.16}
        \arrayrulecolor{black!75}
        \begin{tabularx}{\linewidth}{>{\raggedright\arraybackslash}X|r@{\hspace{5pt}}r@{\hspace{5pt}}r}
            \noalign{\hrule height 0.9pt}
            \rowcolor{resultsheader}
            \multicolumn{1}{c|}{Variant} & BU $\uparrow$ & UA $\uparrow$ & ASR $\downarrow$ \\
            \hline\hline
            w/o Task-use Authorization & 59.72\% & 27.08\% & 7.64\% \\
            \rowcolor{resultsstripe}
            w/o Parameter Support Tracing & 56.94\% & 13.89\% & 11.81\% \\
            w/o Source Authorization & 86.81\% & 21.53\% & 62.50\% \\
            \rowcolor{resultsstripe}
            w/o Semantic Support & 84.03\% & 22.92\% & 28.47\% \\
            \hline
            \rowcolor{resultshighlight}
            AuthorityFlow & 82.64\% & 27.08\% & 9.03\% \\
            \noalign{\hrule height 0.9pt}
        \end{tabularx}
    \end{minipage}\hfill
    \begin{minipage}[t]{0.40\linewidth}
        \vspace{0pt}
        \captionof{table}{Auxiliary verifier results.}
        \label{tab:verifier_generalization}
        \centering
        \footnotesize
        \setlength{\tabcolsep}{2pt}
        \renewcommand{\arraystretch}{1.74}
        \arrayrulecolor{black!75}
        \begin{tabularx}{\linewidth}{>{\raggedright\arraybackslash}X|r@{\hspace{5pt}}r@{\hspace{5pt}}r}
            \noalign{\hrule height 0.9pt}
            \rowcolor{resultsheader}
            \multicolumn{1}{c|}{Verifier} & BU $\uparrow$ & UA $\uparrow$ & ASR $\downarrow$ \\
            \hline\hline
            Llama-3.3-70B & 62.50\% & 10.42\% & 24.31\% \\
            \rowcolor{resultsstripe}
            GPT-4o & 75.69\% & 22.92\% & 15.28\% \\            \hline
            Qwen3.6-27B &  82.64\% & 27.08\% & 9.03\% \\
            \noalign{\hrule height 0.9pt}
        \end{tabularx}
    \end{minipage}

    \par\vspace{12pt}
    \begingroup
    \makeatletter
    \long\def\@makecaption#1#2{%
        \vskip\abovecaptionskip
        {\raggedright #1: #2\par}%
        \vskip\belowcaptionskip}
    \makeatother
    \begin{minipage}[t]{0.48\linewidth}
        \vspace{0pt}
        \centering
        \includegraphics[width=\linewidth]{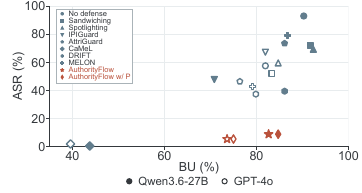}\par
        \nointerlineskip
        \captionof{figure}{Security--utility trade-off on FactAuthBench. Higher BU and lower ASR indicate better performance.}
        \label{fig:security-utility}
    \end{minipage}\hfill
    \begin{minipage}[t]{0.48\linewidth}
        \vspace{0pt}
        \centering
        \includegraphics[width=\linewidth]{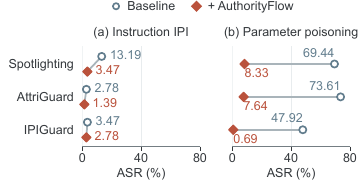}\par
        \nointerlineskip
        \captionof{figure}{Complementarity with IPI defenses. Panel (a) uses vertical offsets for readability.}
        \label{fig:defense-combination}
    \end{minipage}
    \endgroup
\end{figure}

We conduct ablations on Qwen3.6-27B, modifying one design choice at a time while keeping the remaining verification pipeline unchanged. As shown in Table~\ref{tab:ablation}, Task-use Authorization mainly preserves utility: removing it reduces BU from 82.64\% to 59.72\% while leaving ASR nearly unchanged (7.64\%), indicating that task-aware authority avoids over-filtering caused by fixed source-level trust.
Removing Parameter Support Tracing reduces BU/UA to 56.94\%/13.89\% and increases ASR to 11.81\%, showing that direct evidence alone cannot capture parameter dependencies mediated by intermediate facts. Source Authorization and Semantic Support are complementary: removing them increases ASR to 62.50\% and 28.47\%, respectively. Thus, parameter verification requires both determining who is authorized to influence a parameter and whether the resulting value is actually supported.

Table~\ref{tab:verifier_generalization} further evaluates AuthorityFlow with different auxiliary models. AuthorityFlow remains effective with GPT-4o and Llama-3.3-70B as auxiliary models, although their performance degrades to different degrees compared with the default Qwen3.6-27B configuration. This degradation may partly reflect the mismatch between the auxiliary and agent models, together with differences in their reasoning capabilities.
\label{sec:failure-analysis}

\section{Related Work}

\subsection{Indirect Prompt Injection Attacks}
Indirect Prompt Injection (IPI) embeds malicious payloads into external sources accessed during task execution, allowing untrusted content to manipulate model behavior~\citep{pasquini2024neural,chen2025topicattack}. While early work established remote webpages and documents as attack vectors~\citep{greshake2023not}, subsequent studies have expanded both evaluation settings and attack goals~\citep{zhang2025agent,ye2024toolsword,johnson2025dangers}: BIPIA evaluates instructions embedded in external content~\citep{yi2025benchmarking}, InjecAgent targets tool-use agents~\citep{zhan2024injecagent}, and AgentDojo benchmarks dynamic multi-step environments~\citep{debenedetti2024agentdojo}. Recent efforts further investigate adaptive payloads tailored against deployed defenses~\citep{zhan2025adaptive,wang2025agentvigil,nasr2026attacker}.

Despite these developments, parameter-level manipulation and the validity of evidence supporting critical arguments remain underexplored. To bridge this gap, we introduce FactAuthBench for systematic evaluation.

\subsection{Indirect Prompt Injection Defenses}

Existing IPI defenses broadly include training-based robustness, injection detection, and runtime behavioral detection. Training-based methods modify model behavior to resist untrusted instructions~\citep{chen2025secalign,chen2025meta,piet2024jatmo}: Instruction Hierarchy teaches models to respect instruction privilege levels~\citep{wallace2024instruction,wu2025instructional}, while StruQ separates instructions from data and trains models to ignore injected instructions in data regions~\citep{chen2025struq}. Detector-based methods such as DataSentinel~\citep{liu2025datasentinel} and PIGuard~\citep{li2025piguard} instead identify potentially injected inputs before they influence downstream execution~\citep{wen2025defending,chen2025can,shi2025promptarmor}. Runtime defenses infer attacks from behavioral changes: Known-Answer Detection uses predefined probes~\citep{liu2024formalizing}, MELON compares tool calls through masked re-execution~\citep{zhu2025melon}, and AttriGuard uses counterfactual replay to attribute tool behavior to untrusted observations~\citep{he2026attriguard}.

A smaller group of system-level defenses explicitly considers execution and data constraints~\citep{shi2025progent,wu2024isolategpt}. CaMeL extracts control and data flows from trusted user requests and applies capability-based policies~\citep{debenedetti2025defeating}, while DRIFT constructs execution traces and parameter constraints and verifies deviations at runtime~\citep{li2026drift}. However, these approaches do not explicitly model task-specific authority between individual information sources and parameter-relevant facts. AuthorityFlow addresses this gap by constraining which sources may determine task-critical parameters while preserving legitimate multi-source information use.
\label{sec:related-work}

\section{Limitations and Future Work}
\label{sec:limitations}

AuthorityFlow's authority construction relies on contextual information, including user task specifications, tool interfaces, and source provenance. Highly ambiguous user inputs or incomplete tool metadata may therefore introduce uncertainty into the resulting authorization boundaries. At the architectural level, the current design uses an independent verifier model for dual verification. Improving the verifier's robustness against complex adversarial reasoning and reducing its inference overhead remain important practical considerations. Finally, while AuthorityFlow focuses primarily on detecting and intercepting unauthorized parameter-level influence, extending this paradigm to adaptive authority inference and automated task recovery in dynamic environments presents a promising direction for future work.

\section{Conclusion}
\label{sec:conclusion}

In this work, we identify parameter poisoning as a distinct and underexplored attack surface in Indirect Prompt Injection (IPI), where adversaries manipulate task-critical tool arguments without requiring the agent to deviate from its intended actions. To address this threat, we propose AuthorityFlow, a task-aware parameter authorization framework that establishes task-specific authority boundaries, traces factual parameter dependencies, and jointly verifies semantic support and use-specific source authorization. We also introduce FactAuthBench to systematically evaluate agent vulnerability and defense effectiveness against parameter poisoning. Comprehensive experiments demonstrate that AuthorityFlow substantially reduces attack success rates while preserving benign task utility across diverse foundation models. Our findings demonstrate that securing LLM agents against indirect manipulation requires governing not only which instructions they execute, but also which information sources are authorized to determine task-critical parameters.

\bibliography{references}
\bibliographystyle{paper_style}

\end{document}